# THE SEARCH FOR BLACK HOLES IN X–RAY BINARIES: AN UPDATE

L. STELLA*

*Osservatorio Astronomico di Brera, Via Brera 28, I-20121 Milano*

G.L. ISRAEL*

*International School for Advanced Studies (SISSA-ISAS), Via Beirut 2, I-34013 Trieste*

S. MEREGHETTI

*Istituto di Fisica Cosmica, CNR, Via Bassini 15, I-20133 Milano*

and

D. RICCI*

*Dipartimento di Fisica, Università di Roma "La Sapienza", P.le Aldo Moro 2, I-00185 Roma*

ABSTRACT

This paper reviews some of the recent observational results on stellar mass black hole candidates. Over the last decade, much of the progress in this field has been achieved through the study of transient X–ray binary sources which shine only sporadically in the X-ray sky and undergo much larger luminosity variations then most persistent sources. There are currently six stellar-mass black hole candidates for which the mass of the compact object is estimated to be well above the maximum neutron star mass. The three most recently identified black hole candidates of this kind belong to a class of transient X–ray sources with peculiar spectral properties. Several new transients of this class are being discovered and studied every year thanks to more continous monitoring of the X–ray sky. Many of them are likely to contain other black hole candidates. The growing number of candidates and the large luminosity variations (and therefore, accretion rate variations) in transient sources will allow to study the chacteristics of accreting black holes with an unprecedented detail.

## 1. Introduction

A large number of collapsed objects has been discovered and made accessible to detailed study through observations in the X–ray band. Accretion of matter in the strong gravitational field of these objects is, in most cases, responsible for the efficient production of radiative energy. In the case of accreting collapsed objects with stellar mass ($M \leq 20$ M$_\odot$) the bulk of the energy is radiated away in the X–ray band.

X–ray binaries consist of a neutron star or a stellar mass black hole accreting matter from a non–collapsed companion in a close binary orbit. Historically, these systems have played a crucial role in the identification of the first black hole candidate and the measurement of basic physical parameters of compact objects[1] (such as the mass and the magnetic field). There is a growing body of evidence that the central region of active galactic nuclei (AGNs) hosts an accreting black hole with a mass of $\sim 10^5 - 10^8$ M$_\odot$. The observation and measurement of strong gravitational field effects is in principle possible in all these classes of sources, since most of the radiation is produced in the vicinity of the collapsed object. However, a number of largely unknown effects related, e.g., to the dynamics, MHD and radiative–transfer of accretion complicates the interpretation of the observations and only relatively little information on the strong field regions has been obtained so far. Double neutron star binary systems containing a radio pulsar have provided a much *cleaner* laboratory for testing the predictions of gravitational theories. From the study of the propagation delays of the pulsar signal in these systems it has been possible to accurately test general relativity, but only in relatively weak fields[2] ($R \leq 10^{-6} GM/c^2$).

---

*Affiliated to I.C.R.A.

This review concetrates on the recent progress in the search for black holes in X–ray binaries. Over the last decade, X–ray transient sources have proven especially useful for investigating accretion into collapsed objects over a wide range of source luminosities and, therefore, mass accretion rates. A subclass of X–ray transient sources seems to be associated with accreting black hole candidates. The potential of these sources for the identification of new candidates and the study of accretion phenomena close to a black hole is outlined. Some recent results on two persistent X–ray sources, meeting the phenomenological spectral criterion for black holes candidates, are also summarized.

## 2. X–ray binary basics

Luminous X–ray binaries ($L_x > 10^{35}$ erg/s) are often classified as low mass and high mass systems depending on the mass of the donor star. While this classifications leaves unspecified the nature of the accreting collapsed object (which indeed can be a neutron star or a black hole in either class), it allows to separate the phenomenology of the X–ray sources and their optical counterparts in a natural way (see Table 1).

### 2.1. High mass X–ray binaries

High mass X–ray binaries (HMXRBs) contain an early type (OB) star with a mass of $> 5$ M$_\odot$ and have a galactic disk distribution characteristic of young stars (population I). Mass transfer in most of these systems takes place because part of the intense stellar wind emitted by the OB star is captured by the gravitational field of the collapsed object. The energy production budget in HMXRBs is often dominated by the optical luminosity of the OB star, with the X–ray flux emitted in the vicinity of the collapsed object providing only a small perturbation. Correspondingly, the optical spectra are stellar–like.[3]

Periodic X–ray pulsations with periods ranging from $\sim 0.069$ to $\sim 1455$ s are present in a large number (33) of HMXRBs. This signal originates from the beamed radiation which is produced close to the magnetic poles of a young accreting neutron star with a surface field of $\sim 10^{12}$ Gauss. Due to the misalignment of the magnetic and rotation axes, the neutron star rotation modulates in a light–house fashion the X–ray intensity observed from the earth. Period (or phase) changes introduced by the binary motion allow to measure some of the orbital parameters of these systems. Together with the duration of the X–ray eclipse (which is observed in several HMXRBs) and the Doppler velocity and photometric modulations of the optical star, these measurements provide the absolute orbital solution and the masses of the two components. Secular spin period changes arise because of the torque exerted on the neutron star magnetosphere by the accreting matter.[4] X–ray pulsations from luminous X–ray binaries provide an incontrovertible signature of accretion onto a magnetized neutron star.

*2.2. Low mass X–ray binaries*

Low mass X–ray binaries (LMXRBs) typically contain a late type (usually spectral types K or M) low mass donor star (or, in the case of very short period binaries, a white dwarf, WD). About 12 LMXRBs are inside globular clusters and many of the others are concentrated in the vicinity of the galactic bulge, therefore indicating a distribution characteristic of old stars (population II).[5] The short orbital periods of LMXRBs are usually inferred from the orbital modulation of the optical and/or X–ray flux, rather than X–ray eclipses proper[6] (which are rare) . In most cases mass transfer takes place because the low mass companion overfills the critical effective potential surface (the Roche lobe) and spills matter with high specific angular momentum toward the collapsed star, causing the formation of an accretion disk. The intrinsic optical luminosity of the low mass companion is orders of magnitude lower than the X–ray luminosity emitted by the accreting collapsed object. The spectral features of the late type companion are usually outshined by the reprocessing at optical wavelengths of the X–ray flux intercepted by the accretion disk and the star.

X–ray pulsations have been found only in a very small number of LMXRBs. Much more frequent, instead, is the phenomenon of (type I) X–ray bursts, sudden rises of X–ray luminosity which typically last for tens of seconds, show a characteristic cooling in the decay phase and recur on timescales of hours. These bursts account for only a small fraction of the time–averaged luminosity of LMXRBs. They originate from thermonuclear flashes in the freshly accreted matter on the surface of a neutron star. Therefore, type I X–ray bursts provide another clear signature of accretion onto a neutron star.

In a few LMXRBs the bursting activity ceases[7] when the persistent emission X–ray luminosity increases above a level of $\sim 10^{37}$ erg/s. The X–ray properties of this state resemble those of a number of high–luminosity LMXRBs ($L_x \geq 10^{37}$ erg/s), such as Sco X–1, Cyg X–2 and many others which populate the galactic bulge. It is thus inferred that also the latter systems contain accreting neutron stars.

*2.3. X–ray pulsations versus type I X–ray bursts*

X–ray pulsations and bursts are mutually exclusive properties of X–ray binaries: type I bursts have never been observed from an X–ray pulsar and no coherent pulsations have ever been detected from a type I burst source.

Pulsations are not expected from the old neutron stars in bursting sources if their magnetic field has decayed to $\sim 10^8$ Gauss, a value below which accretion is not significantly funneled close to the magnetic poles. On the other hand, LMXRBs are likely progenitors of the old, *recycled* millisecond pulsar which are found in increasing numbers especially in globular clusters.[8] In this case the neutron star magnetic field of $\sim 10^9 - 10^{10}$ Gauss inferred from the radio pulsar observations should also characterize the LMXRB stage, and low amplitude X–ray pulsations in the millisecond range would be expected if accretion occurs preferentially along the magnetic field lines. The search for fast pulsations in LMXRBs still continues.

Table 1: Classification of X-ray binaries

| Properties | HMXRBs | LMXRBs |
|---|---|---|
| donor star | O–B ($M > 5$ M$_\odot$) | K–M or WD ($M < 1$ M$_\odot$) |
| population | I | II |
| $L_x/L_{opt}$ | 0.001 - 10 | 100 - 1000 |
| optical spectrum | stellar–like | reprocessing |
| X–ray spectrum | usually hard | usually soft |
| orbital period | 1 - 100 d | 10 min - 10 d |
| X–ray eclipses | common | rare |
| X–ray pulsations | common (0.1–1000s) | rare (1–100s) |
| type I X–ray bursts | absent | common |
| X–ray QPOs | rare (0.001–1Hz) | common (1–100Hz) |
| collapsed object | neutron star or black hole | neutron star or black hole |

Viceversa, type I bursts might not occur in X–ray pulsar binaries because the strong magnetic fields ($\sim 10^{12}$ Gauss) of young neutron stars confines the infalling plasma to the polar caps, therefore dramatically increasing the accretion rate per unit area (compared to weakly magnetic neutron stars) and giving rise to steady (as opposed to flash–like) thermonuclear burning in the accreting material.[9,10]

While pulsations and type I X–ray bursts are clearly explained and imply the presence of an accreting neutron star, a variety of other phenomena characteristic of X–ray binaries still awaits for unambiguous interpretation. These include aperiodic variability (shot, red, very–low frequency, low–frequency, high–frequency noises), quasi–periodic oscillations (QPOs), some spectral and/or activity states, and continuum and discrete spectral components. At a phenomenological level, a number of regularities and correlations have emerged which provide the basis for further classification and study.

## 3. Persistent versus transient X–ray binaries

Besides persistent sources the X–ray sky is populated by a number of transient sources which remain in their quiescent state for most of the time and sporadically undergo bright outbursts with peak luminosities of $10^{36} - 10^{38}$erg/s , durations ranging from weeks to months, and recurrence timescales of 1–10 years or longer.

Throughout the years a very clear analogy of the X–ray characteristics of bright transient sources with those of persistent sources has emerged. In particular, a number of X–ray transients display coherent X–ray pulsations or bursts, testifying to the presence of neutron stars, which undergo sporadic surges of accretion. Like in the case of persistent sources, X–ray bursting transients have low mass companions and relatively soft X–ray spectra, whereas X–ray pulsations are usually observed from transients in Be–star high mass binaries which are characterized by hard X–ray spectra[11] extending up to several tens of keV.

The identification of the optical counterparts of bright transients, in crowded and often heavily absorbed regions of the galactic plane, is often made easier by the optical flux increase associated with the outburst. In the case of low mass systems, in particular, the reprocessing of high energy radiation can induce an increase of more than a factor of 100 above the quiescent optical flux level.[5] Contrary to the case of persistent low mass X–ray binaries, detailed photometric and spectroscopic studies of the companion star are often possible in low mass transient sources, due to the fact that in the quiescent state its optical spectrum is not dominated by the reprocessing of X–ray radiation or by the emission from the accretion disk around the collapsed object.

X–ray transients sources are also extremely useful in that they allow to investigate accretion onto collapsed stars over a much larger range of X–ray luminosities, and therefore, accretion rates, than persistent sources.

## 4. Black hole candidates in X–ray binaries

Despite the increasing evidence in favor of very massive black holes ($10^6 - 10^9$ M$_\odot$) in AGNs, black hole candidates in X–ray binaries provide still the strongest case for the existence of completely collapsed objects, characterized by an event horizon.

### 4.1. The mass criterion

The chain of arguments that leads to the identification of a black hole candidate in an X–ray binary through the *mass criterion* can be summarized as follows:[12,13]

(i) The luminosity of the X–ray source is high ($> 10^{36}$ erg/s) and possibly characterized by fast aperiodic variability ($< 1$ s): therefore, the binary must contain an accreting compact object.

(ii) No pulsations or type I X–ray bursts are present.

(iii) Optical spectro–photometric observations allow to determine the orbital period and the Doppler velocity modulation of the donor star around the collapsed object. The optical mass function of the system, $f_{opt}(M) = (M_x \sin i)^3/(M_x + M_{opt})^2 = P_{orb} K^3/2\pi G$, is thus determined, where $K = v \sin i (1 - e^2)^{1/2}$ and $v \sin i$ is the semi–amplitude of the Doppler modulation.

(iv) The mass of the optical star and the inclination are measured, or at least constrained, based on the distance, the optical spectrum and luminosity of the optical star, its size relative to the Roche–lobe and arguments related to the lack of

X–ray eclipses and the presence of optical polarization. By using $f_{opt}(M)$, the mass of the compact object, $M_x$, is thus measured or constrained.

(v) If the mass of the compact object is determined to be larger then the canonical maximum mass of a neutron star predicted by the theory, $M_x > M_{max} \simeq 3.2$ $M_\odot$,[14,15] then the conclusion is reached that the accreting compact object is most likely completely collapsed and is, therefore, a *black hole candidate*.

The identification of the first black hole candidate in the early seventies, Cyg X–1, represents one of the most important achievements of X–ray astronomy.

### 4.2. Cyg X–1

Cyg X–1 is a highly variable and luminous HMXRB ($L_x \sim 10^{37} - 10^{38}$ erg/s), with an orbital period of 5.6 days an optical mass function of $f_{opt}(M) \simeq 0.25$ $M_\odot$. The most likely values of the companion mass, $M_{opt} \simeq 30$ $M_\odot$, and the inclination, $i \simeq 30°$, provide an estimate of the mass of the compact object of $M_x \sim 16$ $M_\odot$.[16]

It is important to realize that the value of the optical mass function provides an absolute lower limit to the mass of the compact object, corresponding to the unrealistic situation in which $M_{opt} = 0$ and $i = 90°$. In the case of Cyg X-1, however, $f_{opt}(M)$ is only $\sim 0.25$ $M_\odot$ and deriving a reliable upper limit to $M_x$ is a difficult task. In a number of studies a devil's advocate approach was adopted in order to decrease the estimated value of $M_x$ below 3.2 $M_\odot$. This proved very useful in verifying the validity of the assumptions underlying the mass estimate. Indeed the velocity curve of the companion star and therefore, the mass function, might be affected by several sources of systematic uncertainties. These include tidal distortons, non–synchronous rotation and the effects of X–ray heating of the companion star, as well as contamination by emission lines from an accretion disk, gas streams or the companion's wind.[12] Moreover, the value of $M_{opt}$ determined from the companion's spectrum might be underestimated by a factor of 2 or more (in those X–ray pulsar HMXRBs for which a dynamical determination of the masses is available, the companion star is often undermassive for its spectral class). To circumvent this difficulty a different method was devised which sets a lower limit to the black hole mass as function of the distance, based only on the mass function, the lack of X–ray eclipses and the dereddened flux of the companion star.[17]

The most conservative assumptions lead to the conclusion that in Cyg X–1 $M_x > 3$ $M_\odot$. However, $M_x$ measures the total mass around which the supergiant star orbits. In a system like Cyg X–1, there would be room to accomodate a neutron star and, for example, a 8 $M_\odot$ B star in a tight orbit, which in turn both revolve around the O supergiant. The spectrum of such a B star would be difficult to detect, while $M_x$ would be compatible with the measured value. Such a triple star model, which clearly does not require the presence of a black hole, is still marginally viable. Despite this caveat, Cyg X–1 remained the most reliable black hole candidate for more than a decade.

The compact object in most HMXRBs has been determined to be a pulsating neutron star. Detailed studies of the often heavily reddened optical counterparts of X–ray binaries close to the galactic plane have proven difficult. For persistent LMXRBs the problem is even more pronounced, because of the intrinsically low optical luminosity and the absence (or near absence) of spectral features from the companion low mass star. Based on the X–ray characteristics of Cyg X–1, two *phenomenological signatures* of accreting black holes were suggested. These were aimed at selecting *tentative black hole candidates* among X–ray binaries, the optical counterparts of which could be studied in greater depth in order to apply the mass criterion.

Figure 1: 1–40 keV X-ray spectra of the HMXRB black hole candidates Cyg X–1, LMC X–3 and LMC X–1. The high and low state spectra of Cyg X–1 are indicated. [18]

*4.3. Time variability criterion*

The X–ray flux of Cyg X–1 shows a pronounced aperiodic flickering on timescales of the order of $\sim 1$ s or less. Significant structure has been detected down to a few milliseconds. These variations were modelled in terms of a simple shot noise model as early as 1972 and were suggested to arise by some form of instability in the accreting matter in the vicinity of the black hole.[16]

In the early eighties, a transient source, V0332+53, was observed, which together with $\sim 4.4$ s periodic pulsations, displayed pronounced rapid aperiodic variations similar to those of Cyg X–1.[19] This provided the first clear counterexample of the time variability criterion by showing that the accreting magnetic neutron star in an X–ray pulsar system could also produce fast aperiodic flickering. By now, virtually all classes of accreting compact objects in X–ray binaries are known to display rapid aperiodic variations somewhat similar to those observed from Cyg X–1.[20,21] More detailed investigations are required to characterize the time variability of black hole candidates and revise the criterion accordingly.

### 4.4. Spectral criterion

Cyg X–1 displays different spectral states. During most of the time the source is in the *low state*, characterized by a power–law X–ray spectrum with a logarithmic slope of about −1, which steepens above ∼ 100 keV and extends up to energies of several hundred keV. The X–ray luminosity is ∼ $10^{37}$ erg/s. In the *high state* the luminosity increases by a factor of ∼ 5 − 10 due to the presence of an additional spectral component, shortwards of ∼ 10 keV.[16] The high state spectrum of Cyg X–1 is, therefore, much softer than the low state spectrum (Fig. 1).

The presence of spectral states similar to those of Cyg X–1 and/or a high energy tail extending to hundreds of keV has also been used as phenomenological signature of accretion onto black holes. As described in Section 4.6, the application of this criterion has been very successful over the last few years. However, there are several exceptions. Even out of the first two tentative black hole candidates that were suggested on the basis of the time variability and spectral signatures, only GX 339–4 might contain a black hole.[22,23] Type I X–ray bursts were instead observed from Cir X–1, which proved that the system contains an accreting neutron star.[24] This emphasises that, despite its success, the spectral criterion is not fully reliable.

### 4.5. Black hole candidates in the LMC and X–ray colours

Two other black hole candidates were identified in the early eighties from detailed optical observations of two HMXRBs in the Large Magellanic Cloud. The likely mass of the compact object in LMC X–3 and LMC X–1 was determined to be ∼ 9 $M_\odot$ and ∼ 6 $M_\odot$, respectively.[25,26] The value of the optical mass function, though relatively large for LMC X–3 (∼ 2.3 $M_\odot$), does not, in itself, exclude the presence of a neutron star in either of the two systems (see Table 2). Rather, systematic uncertainties in the luminosity and orbital velocity of the optical star in LMC X–3[27] and the relatively large error in the mass function of LMC X–1 do not allow to completely rule out unlikely scenarios in which the mass of the compact object is < 3 $M_\odot$. In this sense LMC X–3 and especially, LMC X–1 provide less reliable black hole candidates than Cyg X–1. The X–ray spectra of LMC X–1 and LMC X–3 are extremely soft and resemble the high state spectrum of Cyg X–1 (Fig. 1).

When an X–ray colour–colour diagram was assembled from observations in the 1.5 − 30 keV band (Fig. 2), it was noticed that the two LMC black hole candidates and Cyg X–1 and GX 339–4 in their high state lie in the left part of the diagram, characteristic of *ultrasoft* spectra.[28] On the contrary, the low state colours of Cyg X–1 and GX 339–4 are harder and close to those of LMXRBs containing an old accreting neutron star. Should the colour–colour diagram be calculated for higher energies (from tens to hundreds of keV, where the observations are still sparse), then the high energy tail would probably make the hardness ratios of black hole candidates the highest (see Fig. 3a). In the 1.5 − 30 keV energy range, instead, X–ray pulsar binaries present with the hardest spectra and tend to occupy the right part of the diagram.

Figure 2: An X-ray colour-colour diagram. Different types of X-ray sources are indicated by different symbols [28].

### 4.6. Black hole candidates in X-ray transients

A number of *ultrasoft* sources, that do not display bursts or pulsations, have colours similar to those of black hole candidates in their high state. Shortly after the optical studies of LMC X–1 and LMC X–3, these sources were suggested as tentative black hole candidates. Among these there is a high incidence of *ultrasoft transient sources*, some of which have been identified as LMXRBs.[11] While in quiescence, these systems provide a rare opportunity to study the companion's optical spectrum and thereby determine the distance and some orbital parameters.

A0620–00 was the first ultrasoft X-ray transient to be studied in quiescence. For two months in 1975 it was the brightest X-ray source in the sky and achieved a peak luminosity of $L_x \sim 10^{38}$ erg/s. Its optical counterpart was identified thanks to the $\sim 6$ mag brightening that accompanied the X-ray outburst. Detailed optical spectro-photometry in quiescence revealed an orbital period of $\sim 7.7$ hr and a low mass K-star companion, whose spectral features were modulated with a velocity semiamplitude of $\simeq 450$ km/s (see Table 2). This provided an optical mass function of $\sim 2.9$ $M_\odot$, almost sufficient in itself to identify A0620–00 as the first black hole candidate in a LMXRB.[30] The likely black hole mass is $M_x \sim 10$ $M_\odot$. A firm lower limit of $M_x > 3.2$ $M_\odot$ is obtained through considerations which are insensitive to the distance and the mass of the companion. This is unlike the black hole candidates in HMXRBs. Moreover, triple star models similar to those suggested for Cyg X–1 would not be viable, because in a compact and optically faint binary like A0620–00 it would be impossible to hide a non–degenerate star of suffinciently high mass (see Fig. 4). A scenario in which two or more degenerate stars (neutron stars and/or white dwarfs) in a very close orbit revolve around the K–star companion can probably be constructed; such a system, however, would be very short lived due to the emission of gravitational radiation or the ejection of one (or more) degenerate star(s). As a black hole candidate, A0620–00 presents, therefore, clear advantages with respect to Cyg X–1; some of these are related to the transient LMXRB character of the system.

Figure 3: (a) Comparison of the wide energy range X-ray spectra from the bright LMXRB Sco X–1 (1), the transient X-ray pulsar HMXRB A0535+26 (2) and the black hole candidate GS 2023+338 [29] (3). Note the spectrum of GS2023+338 extends to much higher energies than the spectra of the other two sources. (b) The 2-40 keV spectral variability of the potential black hole candidate GS2000+25. Note the independent variations of the ultrasoft and power law components [18].

The search for other black hole candidates in transient LMXRBs has been very successful over the last few years. New tentative candidates are being been discovered through the application of the spectral criterion, owing to a more consistent monitoring of the X-ray sky with large field of view detectors. In particular observations in the hard X-ray band ($\geq 30$ keV) have recently allowed to identify several new transient sources, with spectra extending to several hundreds of keV. Discovered and monitored in 1989, GS2023+338 (V404 Cygni) is one such transient; its X-ray spectrum is similar to the low state spectrum of Cyg X–1 (see Fig. 3a). Observations of its optical counterpart have allowed to measure an orbital period of $\sim 6.5$ d and a velocity semiamplitude of $\sim 210$ km/s for the companion star. The very high value of the mass function, $f_{opt} \simeq 6.3$ M$_\odot$, establishes the compact object in GS2023+338 as the best black hole candidate available to date, independent of any consideration concerning the distance, the luminosity, the inclination or the mass of the companion star.[31] It is interesting to note that there is indirect evidence for a third star in a tight orbit around the collapased object in GS 2023+338; the current constraint on the mass of such a star ($M < 0.5$ M$_\odot$) leaves unaffected the conclusion that GS 2023+338 contains a black hole candidate. Another ultrasoft X-ray transient, GS1124-68 (Nova Muscae), has been recently determined to host a black hole candidate[32] based on a mass function of $\sim 3.1$ M$_\odot$.

Table 2: Properties of black hole candidates

| Properties | CygX-1 | LMCX-1 | LMCX-3 | A0620-00 | GS2023+338 | GS1124-68 |
|---|---|---|---|---|---|---|
| class[a] | pers. H | pers. H | pers. H | trans. L | trans. L | trans. L |
| $L_x^{max}$ erg/s | $10^{38}$ | $2 \times 10^{38}$ | $3 \times 10^{38}$ | $1 \times 10^{38}$ | $4 \times 10^{38}$ | $10^{38}$ |
| donor star | O9.7I | late O | B3V | K5V | G/K | K2V |
| d kpc | 2.5 (?) | 55 | 55 | 1 (?) | 2 (?) | 5 (?) |
| V-mag | 9 | 14 | 17 | 18 | 19 | 20 |
| $K$ km/s | $75 \pm 1$ | $68 \pm 8$ | $235 \pm 11$ | $433 \pm 4$ | $211 \pm 4$ | $409 \pm 18$ |
| $P_{orb}$ d | 5.6 | 4.2 | 1.7 | 0.32 | 6.5 | 0.43 |
| $f_{opt}(M)$ M$_\odot$ | $.25 \pm .01$ | $.14 \pm .05$ | $2.3 \pm .3$ | $2.91 \pm .08$ | $6.26 \pm .31$ | $3.1 \pm .4$ |

[a] pers: persistent, trans: transient, H: HMXRB, L: LMXRB,

## 5. Tentative black hole candidates in X-ray binaries

The number of black hole candidates in X-ray binaries has increased from one to six over the last decade. Despite their faintness (often requiring observations with the largest telescopes), the optical counterparts of several other tentative black hole candidates in X-ray transients, in particular the most recently discovered ones, are being studied. Some still remain to be identified. Table 3 provides an up to date list of tentative black candidates, according to the spectral criterion. For the transient sources the year of the outburst is indicated. It is apparent that thanks to a more continuous coverage of the X-ray sky over the last years the number of black hole candidates is rapidly increasing. A0620-00 was observed as an optical nova in 1917; GS2023+338 in 1938 and 1958; some outbursts may have been missed. Recurrence times are, therefore, in the range of tens of years. The total number of transient black hole candidate LMXRBs in the galaxy, though very uncertain, is currently estimated between $\sim 100$ and $\sim 3000$.[33]

Figure 4: Schematic sketch, to scale, of likely models of the HMXRB black hole candidates Cyg X–1 and LMC X-3 and the transient LMXRB black hole candidate A0620–00. The optical companions (shaded regions) are assumed to fill their Roche lobe [13].

Table 3 comprises also several persistent X–ray binaries with an ultrasoft spectrum. In the following subsections we summarize some of recents results on two persistent sources, 4U1957+11 and 4U0142+61: while the former still provides a viable black hole candidate, the X–ray flux from the latter has been shown to result from two X–ray pulsars systems.

### 5.1. *4U0142+61*

In the X–ray colour–colour diagram of Fig. 2, the ultrasoft persistent X–ray source 4U0142+61 occupies the same region of black hole candidates in their "high state". 4U0142+61 lies in the galactic plane (l=129°.4, b=–0°.4) and, despite its small error circle (a few arcseconds), no optical or radio counterparts have yet been identified.[54] During one observation of this source, an additional spectral component was detected above 3 keV within the ∼90 arcmin collimator response of the EXOSAT Medium Energy detector. The overall shape of the spectrum of 4U 0142+61, therefore, held a clear resemblance to the two–component spectra of several black hole candidates. However, ∼25 min periodic oscillations were discovered in the hard spectral component (3–10 keV[54]), leading to the suggestion that this spectral component originated in a different source within the (non–imaging) instrument field of view.[55]

A reanalysis of the EXOSAT data revealed the presence of a coherent modulation with a period of ∼ 8.7 s in the 1–3 keV energy range, where the ultrasoft component from 4U0142+61 dominates.[56] The 4–11 keV light curves did not show any evidence for these pulsations. On the contrary, the 25 min modulation was clearly visible in the 4–11 keV, but not in the 1–3 keV light curves.[57] Even in the absence of a clear detection with an X–ray imaging instrument, the conclusion was reached that the ~ 8.7 s and ~ 25 min pulsations arise from two separate X–ray sources, both containing rotating magnetized neutron stars. This interpretation has been recently confirmed through the detection of both periodicities in an imaging ROSAT PSPC observation, in which the two sources 24 arcmin apart are well resolved.[58]

Table 3: Tentative black hole candidates

| Source | Notes |
| --- | --- |
| GX339-4 | V= 15-21, $P_{orb}$=14.8 hr, pers., bimodal spectral states[34] |
| 4U1755-33 | V= 18, pers., ultrasoft[35] |
| 4U1957+11 | V= 19, $P_{orb}$=9.3 hr, pers., ultrasoft, hard comp.[60] |
| GRS1758-258 | V=no countepart, radio jets, pers., very hard, soft comp.?[36] |
| 1E1740-2942 | no counterpart, pers., very hard, 511 keV line[37,38] |
| SS 433 | V=14, $P_{orb}$=13 days, pers., radio/X jets[39] |
| 4U1543-47 | V=15-17, trans. 1971/1983, ultrasoft[40] |
| 4U1630-47 | no counterpart, recurrent trans.? ultrasoft, hard comp.[41] |
| H1705-250 | V=16-21, trans. 1977, ultrasoft[42] |
| H1743-322 | no counterpart, trans. 1977, ultrasoft[43] |
| A1524-361 | V=15-21, trans. 1974, ultrasoft[44] |
| EXO1846-031 | no counterpart, trans. 1985, ultrasoft, hard comp.[45] |
| GS1354-645 | V=17-22, trans. 1987 (and others if = Cen X-2), ultrasoft[46] |
| GS1826-24 | no counterpart, trans. 1988, hard comp.[18] |
| GS2000+25 | V=17-21, $P_{orb}$=8.3 hr, trans. 1988, ultrasoft, hard comp.[47,48] |
| GRS1009-45 | V=15 during outburst, trans. 1993, very hard, soft comp.[49] |
| GRO J0422+32 | V=13-20, $P_{orb}$=5.1 hr, trans. 1992, very hard[50] |
| GRS 1915+105 | V=no counterpart, IR K=13-14, trans. 1992, very hard, superluminar expans.[51] |
| GRO J1655-40 | V=14-18, trans. 1994, very hard[52,53] |

The two component spectrum revealed by EXOSAT, therefore, resulted from the sum of the spectra of two nearby neutron star systems, rather than from a single black hole candidate. 4U 0142+61 is probably a rare example of LMXRB showing periodic X-ray pulsations, while the source responsible for the 25 minutes periodicity is the Be star X-ray transient RX J0146.9+6121 containing the most slowly rotating known neutron star. Independent of the high energy component, we note that the detection of coherent pulsations from the 4U 0142+61 further weakens the phenomenological criterion that the presence of an ultrasoft X-ray spectral component is characteristic of accreting black holes.

## 5.2. 4U1957+11

4U1957+11 is a relatively poorly studied low mass X–ray binary close to the galactic plane, with an X–ray luminosity of a few $\times 10^{36}$ erg/s. In the X–ray colour–colour diagram it lies close to the region occupied by black hole candidates in their high state (see Fig. 2). This suggests that 4U1957+11 might host a black hole candidate.

EXOSAT observed 4U1957+11 twice: in 1983 and 1985. The main component of the *EXOSAT* spectra is consistent with a power law model with exponential cutoff characterized by a spectral index of $\sim 0.2$ and a cutoff energy of $\sim 2$ keV for both the 1983 and 1985 observations. This spectral form provides a simple approximation to the thermal Comptonization spectrum of Sunyaev and Titarchuk (1980). Thermal Comptonization models (and their approximations) are found to provide very good fits to the main spectral component of a number of high state black hole candidates and LMXRBs containing old weakly magnetic neutron stars.[59] In high state black hole candidates this spectral component is somewhat softer then that of 4U1957+11 ($E_{\rm cut} \sim 1.4$ keV in LMC X–3 and LMC X–1). In LMXRBs containing an old neutron star the component is instead harder (e.g. $E_{\rm cut} \sim 7$ keV in 4U1705–44 and $E_{\rm cut} \sim 5$ keV in 4U1636–53). A classification of the compact object in 4U1957+11 based on the characteristic of its Comptonized spectral component is, therefore, premature.[60]

On the other hand the 1985 *EXOSAT* spectra and, especially, the Ginga spectra[61] show the presence of a high energy power–law component above energies of 10 keV. This kind of two–component spectra has been observed from several black hole candidats in their high state, the luminosity of which, however, is a factor of $> 10$ higher than that of 4U1957+11.

A Fe K–shell emission feature is detected at an energy of $7.06 \pm 0.09$ keV in the 1985 data. The centroid energy of such a line is consistent only with $K\alpha$ transitions from H–like ions ($E_0 = 6.96$ keV) and is among the highest revealed with the *EXOSAT* GSPC.[62] $K\alpha$ transitions from any of the lower ionization stages of iron require a substantial blueshift. Similar to the modelling of the Fe K–lines from other X–ray binaries and AGNs, such a blueshift could result from (relativistic) Doppler effects due to bulk plasma motions in the vicinity of the collapsed object (see also Section 6.g). In particular, the observed feature might correspond to the blue horn of the characteristic profile produced in the innermost region of a relativistic accretion disk.[63] 4U1957+11 show only a moderate variability on timescales of a few hours, similar to black hole candidates in their high state (e.g. LMC X–3[64,65]). A detailed search for periodicities revealed no coherent modulation. 4U1957+11 still provides a potentially viable black hole candidate, despite the somewhat low X–ray luminosity and anomalous ultrasoft component.

## 6. Open issues in the physics of black hole accretion

The rapidly growing number of candidates, together with the large range of X–ray luminosities and mass accretion rates that can be studied in transient systems, will allow to address in greater detail a number of open issues related to the physics of accreting black holes. Among these:

(a) What is the origin of the ultrasoft spectral component, characteristic of the *high state*, and the power law component, which is dominant in the *low state* ? The intensity of each of the two components can vary independently of the other; the tentative black hole candidate GS2000+25 provides the best example of this behaviour (Fig. 3b[66,65]). Accretion disk models have been developed which comprise a *cold* disk emitting ultrasoft X–rays and a *hot* phase, in the geometry of either a corona above the accretion disk or a central bulge, which produces the power law spectrum.[67-69] The ultrasoft X–ray component might originate from the superposition of black body spectra (possibly modified by the effects of electron scattering) or from Comptonization by low energy thermal electrons.[64,59,65] The high energy power law component is likely produced by Comptonization. The latter process may work either by upscattering the soft photons produced in the cold disk, or by downscattering the $\gamma$–ray photons produced in a $e^+ - e^-$ pair dominated plasma.[70]

(b) Why is substantial emission above a few hundred keV characteristic of accretion into black holes ? Only very preliminary explanations are available. Possibly the absence of a star surface and of the associated emission of soft photons prevents cooling of the Comptonising region, and higher electron energies can be mantained than in the case of accreting neutron stars. Recent observations indicate that also the persistent emission of X–ray burst source possess a high energy tail extending to 100–200 keV[71] There is, however, a preliminary indication that in these sources the power law slope of the high energy spectrum is steeper than that of black hole candidates.

(c) What is the origin of the emission feature at energies of 400–600 keV or higher observed in some black hole candidates ? In the case of Cyg X–1 a variable broad bump at energies of 400–1500 keV has been observed, which could correspond to the Wien peak of a Comptonized spectrum or, alternatively, to a blueshifted $e^+ - e^-$ annihilation line.[72] A much narrower feature around $\sim$ 500 keV has been recently reported from the ultrasoft X–ray transient GS 1124–68 which is strongly suggestive of an $e^+ - e^-$ annihilation line.[73,74] Also the high energy source 1E1740.7–2942 near the galactic center displayed a variable bump at 300–600 keV, probably related to annihilation processes in relatively cold electron–positron plasma.[75,37] Further observations with higher spectral resolution and a more extensive coverage of a number of sources are required to confirm this interpretation. Annihilation lines might provide an important dignostic of pair–dominated plasmas.

(d) Is there cold optically thick matter in the vicinity of accreting black holes ? The low state spectrum of Cyg X–1 and, especially, the variable absorption spectra GS2023+33 display a broad excess at $\sim 10 - 40$ keV, above a simple power law spectrum.[65,66] A similar excess is observed in the X–ray spectrum of a number of Seyfert I galaxies.[76] This feature is best interpreted in terms of reflection by relatively cold optically thick matter in the vicinity of the collapsed object. Reflection is most effective above $\sim$ 10 keV, where the photoelectric absorption by heavy ele-ments is small, and below $\sim 300$ keV, where the effects of Compton recoil are not dominant.[77] The study of this spectral component can provide insights about the physical state and geometry of matter in the vicinity of the black hole.

(e) Is there a type of short term X–ray variability which is characteristic of black hole accretion ? While virtually all classes of accreting compact stars are by now known to produce fast aperiodic variations ($\tau < 1$ s), more detailed studies based on power spectral and cross–spectral analyses suggest that there might exist a type of aperiodic variability which is specific to accretion into black hole candidates.[78,21] This would also allow revision of the time variability criterion for identifying tentative black hole candidates

(f) Is there a connection between the QPOs in the X–ray flux of black hole candidates and those from accreting neutron star systems ? QPOs have already been revealed from several black hole candidates (and tentative black hole candidates). The QPO frequency is relatively low in the HMXRB black hole candidates LMC X–1 ($\sim 0.07$ Hz[79]) and Cyg X–1 ($\sim 0.04$ Hz[80,81]). The QPOs from GX339–4 display different modes, associated to different time variability and spectral states, which hold some interesting analogies with the QPOs observed from LMXRBs containing an old neutron star.[82] The frequency of the QPOs from the ultrasoft transient black hole candidate GS1124–68 is also quite high ($\sim 10$ Hz). Models involving rotating magnetospheres are clearly not applicable to black hole candidate QPOs. Disk instability models seem to be more promising.[83]

(g) Are the Fe K–shell lines produced in the innermost regions of the accretion disks around black hole candidates ? The iron emission features observed at $\sim 6-7$ keV in virtually all classes of accreting collapsed stars, probably provide the most promising diagnostic of the physical conditions and possibly, the dynamics of the innermost regions of accretion flows towards collapsed objects. These lines have been studied in most cases only with modest spectral resolution detectors ($E/\Delta E \leq 10$), which could not resolve the profile of the line. Yet two very important conclusions have been reached. Firstly, the line centroid energy measured from Cyg X–1 ($\sim 6.2$ keV, see Fig. 7a) and the ultrasoft transient 4U1543–47 ($\sim 5.9$ keV) requires a substantial redshift as the rest line energy ranges from 6.4 keV for fluorescence from the lowest ionizations stages of iron, to 6.96 keV for recombination into hydrogenic iron ions.[84,40] Secondly, the iron K$\alpha$ lines are broadened to widths of 1 keV or more for many accreting collapsed objects, such as old neutron stars in LMXRBs and a few black hole candidates.[85,86] Blending of lines from different ionization stages of iron can produce a maximum width of only $\sim 0.5$ keV. The observed line redshifts and widths are unlikely to be caused by Compton scattering of line photons in an electron cloud.[87] The interpretation in which the broadening and shifting of the iron K$\alpha$ lines arises because of the Doppler, transverse and gravitational shifts from high velocity plasma motions in the vicinity of the collapsed object appears to be more natural. The accretion disk modelling of the line profiles requires a general relativistic treatment, as the observed widths and redshifts suggest a line emitting region in the range of tens of Schwarzschild radii.[63] The profiles calculated for a

Keplerian disk orbiting a non–rotating black hole display a characteristic double–horned profiles. Relativistic effects (aberration, time dilation and redshift) combine to produce a blue horn brighter then the red horn, whereas the gravitational and transverse effects can cause an overall redshift of the line. The solid state detectors (resolution of $E/\Delta E \sim 50$) onboard ASCA should provide important new information on the iron K$\alpha$ lines from a number of accreting collapsed stars and possibly allow to resolve their profile. The observation of the characteristic double–horned profile would not only confirm the accretion disk scenario, but would also provide an unprecedentedly sensitive probe of regions as close as a few tens of Schwarzschild radii to the central object. In particular the dimensions (in units of Schwarzschild radii), inclination and emissivity distribution can be measured directly from the profile.[63,88–90]

(h) What causes the ejection of matter at relativistic speeds from stellar mass black holes ? And what is the connection with similar phenomena observed in radio galaxies and quasars ? The evidence for plasma acceleration in galactic compact objects comes from the observation of strong, non–thermal radio outbursts in several sources.[91] The most spectacular examples have been recently found in hard X–ray sources. The source 1E 1740.7–2942,[37] which is also the likely origin of the variable electron/positron annihilation line observed from the galactic center direction,[92] has two oppositely directed radio jets. The jets emanate from a variable, radio core source and terminate with bright lobes which give to the source the appearence of a miniature radio quasar.[38] The recent discovery of superluminal motion in the radio counterpart of the hard X–ray source GRS 1915+105 is even more spectacular.[51] This is the first time that such an effect is observed in an object located within our Galaxy. This object will probably offer the best opportunity in the near future to gain a deeper understanding of relativistic jets.

## 7. Conclusions

Over the last few years a substantial progress has been achieved in the identification and study of black candidates in X-ray binaries, especially through transient systems. For three transients a reliable lower limit to the mass of the collapsed object has been obtained from dynamical studies of the companion star in the quiescent state, when the reprocessing of X-ray radiation does not dominate the emission at optical wavelenghts. The optical counterparts of a number of tentative black hole candidates in X–ray transients remain to be identified and studied; moreover, new transients are being discovered with large field of view X–ray instruments. Therefore, the number of candidates is likely to increase rapidly in the near future.

Extensive X–ray monitoring of the outbursts, with improved spectral resolution and extended energy range, will allow to investigate in detail the properties of accreting black hole candidates over the large luminosity variations which are characteristic of X–ray transients. These studies will bring the level of our understanding of accreting black holes closer to that of accreting magnetic neutron stars.

The mass argument is currently central to the identification of black hole candidates. While based on a small number of very likely hypothesis, the argument provides only an indirect proof of the existence of black holes. Even within the framework of general relativity, compact stellar configurations substantially more massive than neutron stars could be made of, e.g., unknown light and stable fermions, or high–density nongravitationally bound states of baryon matter (Q–stars[93]). A new perspective for the observation and study of the properties of black hole candidates is clearly emerging, which will allow to probe regions closer and closer to the collapsed object. Identifying reliable *strong field signatures* of accreting black holes will require new synergetic efforts of theorists and observers. Ultimately, incontrovertible evidence should be found for the existence of an event horizon.